\documentstyle[12pt, aasms4, graphics]{article}

\begin{document}

\title{REDSHIFTS AND LUMINOSITIES FOR 112 GAMMA RAY BURSTS}

\author{Bradley E. Schaefer$^{a}$, Ming Deng$^{b}$, and David 
L. Band$^{c}$ \\
$^{a}$University of Texas, Department of Astronomy, C-1400, Austin TX
78712 \\
$^{b}$Yale University, Physics Department, P. O. Box 208121, New 
Haven CT 06520-8121 \\
$^{c}$Los Alamos National Laboratory, X-2, MS B220, Los Alamos 
NM 87545}

%This is single spacing
\baselineskip 12pt
%This is double spacing
%\baselineskip 24pt

\begin{abstract}

Two different luminosity indicators have recently been proposed for Gamma
Ray Bursts that use gamma-ray observations alone.  They relate the burst
luminosity ($L$) with the time lag between peaks in hard and soft energies
($\tau_{lag}$), and the spikiness or variability of the burst's light curve
($V$).
These relations are currently justified and calibrated with only 6 or 7
bursts with known red shifts.  We have examined BATSE data for $\tau_{lag}$ 
and V for 112 bursts.  (1) A strong correlation between $\tau_{lag}$ and V 
exists, and it is exactly as predicted from the two proposed relations.  This 
is proof that {\em both} luminosity indicators are 
reliable.  (2) GRB830801 is the all-time brightest burst, yet with a small
V and a large $\tau_{lag}$, and hence
is likely the closest known event being perhaps as close as 3.2 Mpc.  (3)
We have combined the luminosities as derived from both indicators as a
means to improve the statistical and systematic accuracy when compared with
the accuracy from either method alone.  The result is a list of 112
bursts with good luminosities and hence red shifts.  (4) The burst averaged
hardness ratio rises strongly with the luminosity of the burst.  (5) The
burst luminosity function is a broken power law, with the break at 
$L = 2 \times 10^{52} erg$.  The numbers in logarithmic bins scale as 
$L^{-2.8 \pm 0.2}$ above the break and as $L^{-1.7 \pm 0.1}$ below the
break.  (6) The number 
density of GRBs varies with red shift roughly as $(1+z)^{2.5 \pm 0.3}$
between $0.2<z<5$.  This demonstrates that the burst rate follows the star
formation rate at low red shifts, as expected since long bursts are
generated by very massive stars.  Excitingly, this result also provides a
measure of the star formation rate out to $z \sim 5$ with no effects from
reddening, and the rate is rising uniformly for red shifts above 2.

\end{abstract}


\clearpage

\section{Introduction}

One of the most fundamental questions in both astronomy and Gamma-Ray
Burst (GRB) research is always the distance to sources.  From their
discovery in 1973 until 1997, the distance scale to GRBs was uncertain by
11 orders of magnitude.  Since 1997, the discovery of low-energy counterparts
(Costa et al. 1997, van Paradijs et al. 1997, Frail et al. 1998) has lead
to the measurements of red shifts of GRBs (Metzger et al. 1997), thus
proving
that most GRBs are at cosmological distances.  Nevertheless, just over a
dozen
bursts currently have known red shifts and this small sample does not
allow detailed demographic studies.
  
At the Fifth Huntsville Gamma-Ray Burst Symposium in October 1999, two
research groups announced the discovery of two different GRB luminosity
indicators wherein the luminosities and distances could be derived from
gamma-ray data alone.  The first indicator relates the luminosity to the
lag, which is the time delay between the peaks for light curves of
energies roughly 25-50 keV and 100-300 keV (Norris, Marani, \& Bonnell
2000).  The second indicator relates the luminosity to the variability,
which is the variance of the light curve around a smoothed light curve
(Fenimore \& Ramirez-Ruiz 2000).  High luminosity bursts have near-zero
lags and spiky light curves, while low luminosity GRBs have long lags and
smooth light curves.  Both relations were calibrated with only 6 or 7
bursts with known red shifts, so it is problematic whether the claimed
relations are fortuitous since a small number of random points can easily
look like a straight line on log-log plots.

The discovery of luminosity indicators that only use gamma-ray data opens
the possibility of using the entire BATSE database for demographic work,
without having to await the accumulation of optical red shifts.
Unfortunately, it might be several years before enough optical red shifts
are found to provide independent confirmation of the validity of the
luminosity indicators.

In this paper, we present a means of proving {\em both} luminosity indicators
without having to measure any additional red shifts.  The idea is that the
existence
of a lag/luminosity relation and a variability/luminosity relation
predicts a particular lag/variability relation, and this prediction can be
tested with the BATSE data in hand.  If either one or both of the two
luminosity indicators are not true, then the predicted lag/variability relation
will not be found.  Since the lag/variability relation can be tested for a
large number of bursts independent of the calibration bursts, a successful
prediction gives proof that {\em both} luminosity relations are correct.

\section{LAG/VARIABILITY RELATION}

The two luminosity indicators have been originally calibrated with
different definitions of the luminosity which differ substantially for the
same burst.  For this paper, we need a simple definition of luminosity
that can be readily calculated for many BATSE bursts.  So we take the
luminosity to be 

\begin{equation}
	L = 4 \pi D^2 \cdot P_{256} \cdot <E>.
\end{equation}

Here $D$ is the luminosity distance (for $H_o = 65 km \cdot s^{-1}
\cdot Mpc^{-1}$,
$\Omega = 0.3$, $\Lambda = 0.7$), $P_{256}$ is the BATSE peak flux for the
256 ms time scale from 50-300 keV (in units of $photon \cdot s^{-1}
cm^{-2}$),
and $<E>$ is the average energy of a photon for an $E^{-2}$ spectrum
($1.72
\times 10^{-7} erg \cdot photon^{-1}$).  This formulation includes a
K-correction for an $E^{-2}$ spectrum, as appropriate for average bursts
(Schaefer et al. 1994; 1998).  Throughout this paper, the luminosity is
calculated assuming that the radiation is emitted from the source
isotropically.

Norris, Marani, \& Bonnell (2000) found that the luminosities of six
bursts
(with known optical red shifts) are well correlated as a power law with
the lag for the bursts.  The lag, $\tau_{lag}$, is the time delay 
of the maximum cross correlation between the BATSE energy channels 1 (25-50
keV) and 3 (100-300 keV).  In essence, the lag is the time between the
peaks as viewed with hard and soft photons.  Our fit to the data
(excluding GRB980425) gives a lag/luminosity relation of

\begin{equation}
	L_{lag} = 2.9 \times 10^{51} (\tau_{lag}/0.1s)^{-1.14},
\end{equation}

with an rms scatter of 0.26 in the logarithm of luminosity.  The exponent
has an uncertainty of 0.20.  GRB980425 (the burst associated with SN1998bw 
[Galama et al.
1998]) falls greatly below this relation, although its very low luminosity
($2.0 \times 10^{46} erg \cdot s^{-1}$) is indeed qualitatively indicated
by 
its extremely long lag (4 s).

Fenimore \& Ramirez-Ruiz (2000) found that the luminosities of seven
bursts (with known optical red shifts) are correlated as a power law with
the variability of the burst.  The variability, V, is the normalized
variance of the observed 50-300 keV light curve about a smoothed light
curve.  The smoothing is done with a boxcar window with length equal to
15\% of the burst duration.  Corrections are made for red shift effects
(hence requiring an iterative procedure) and for the Poisson variations of
the light curve.  The best fit power law depends substantially on how
systematic errors are included, how the formally negative V values are
handled, and whether GRB980425 is included.  A typical fit is

\begin{equation}
	L_{var} = 10^{52} (V/0.01)^{2.5},
\end{equation}

with an uncertainty of roughly 1.0 in the exponent and a factor of
a few in the proportionality constant.  This is essentially
the same result as given by Fenimore \& Ramirez-Ruiz (2000), and by Reichart 
et al. (2000) for a subtly improved definition of V.  The rms scatter about 
the above relation is roughly 0.6 in the logarithm of the observed luminosity. 
 Again, GRB980425 falls greatly below this relation, although its very low 
luminosity is qualitatively indicated by its extremely low V.

If both equations 2 and 3 are correct, then we can predict that there
should be a lag/variability relation of

\begin{equation}
	V = 0.0021 \cdot \tau_{lag}^{-0.46}.
\end{equation}

To test this prediction, we have taken variability measurements from Fenimore
\& Ramirez-Ruiz (2000) and lag measurements from Band (1997).  The bursts with
V measurements were selected by brightness ($P_{256} > 1.5 photons \cdot s^{-1}
\cdot cm^{-2}$) and duration ($T_{90} > 20 $s).  The $\tau_{lag}$ measurements
were selected for bursts that were complete for roughly $P_{256} > 3.25 
photons \cdot s^{-1} \cdot cm^{-2}$.  Our lags are quantized
to 0.064 s bins, so that an additional uncertainty of 0.032 s should be
added in quadrature to the scatter about the calibration curve to obtain
the total one-sigma error of the lag.  For bursts with low lags (i.e.,
high luminosity events), this quantization error becomes large.  We have
112 GRBs with both $\tau_{lag}$ and V measures.  These are plotted in
Figure 1.

Figure 1 shows a significant lag/variability correlation.  The logarithms of 
$\tau_{lag}$ and $V$ are correlated with $r = -0.45$, which for 112 data
points corresponds to a probability of $7.6 \times 10^{-7}$ for chance 
occurance.  Figure 1 also shows the predicted relation from eq. 4.  The
observed slope is close to that of the predicted slope, while the
intercept is a factor of two low which is within the uncertainties of
eq. 4.  Thus, both the lag/luminosity and the
variability/luminosity relations have passed a severe test involving 112
bursts independent from the original calibration.

We take this successful prediction as strong proof that {\em both}
luminosity
indicators are valid.  If only one of them is valid while the other is
false, then our observed lag/variability relation must certainly be
different than predicted by eq. 4.  If both
luminosity indicators are false, then it would be a very improbable
coincidence that the existence, slope, and intercept of our
lag/variability relation came out as predicted by eq. 4.   

\section{GRB 830801}

GRB830801 is by far the all-time brightest GRB event known.  With a peak
flux of $3.0
photons \cdot cm^{-2} \cdot s^{-1} \cdot keV^{-1}$ averaged from 50 to 300 
keV, a dead time correction by a factor of 1.9, and a smooth light curve for 
the peak 256 ms time interval (Kuznetsov et al. 1986); the peak flux $P_{256}$ is around $1400 photons \cdot s^{-1} \cdot cm^{-2}$.  For comparison, BATSE's 
brightest burst (GRB930131, the `SuperBowl Burst') only has $P_{256} = 105 
photons \cdot s^{-1} \cdot cm^{-2}$.  

GRB830801 was remarked to have no fast light curve variations beyond the
Poisson noise level.  Indeed, a look at the light curve shows an extremely
smooth event, and the tremendous photon statistics allows this
smoothness to be obvious.  In other words, GRB830801 has a very small V.

GRB830801 has a large lag.  This can be quantified from Figure 4 of Kuznetsov 
et al. (1986), from which the peak times of light curves can be read for 
seven energy bands.  These peak times can be estimated to 0.25
s accuracy and plotted as a function of energy.  A smooth fitted curve
through these points allows for identification of the time of peak at energies 
of 35
keV and 170 keV (mean effective energies corresponding to BATSE energy
channels 1 and 3) as 4.0 s and 1.8 s after the trigger respectively.  The
uncertainty is of order 0.2 s, primarily due to a short extrapolation to
35 keV from the lowest energy GRB830801 light curve (from 39-68 keV).  So
the lag is $2.2 \pm 0.2$ seconds.

Both a very small variability and a very large lag imply that GRB830801
had a very low luminosity.  If we use eq. 2, then $L_{lag} = 8.5 \times 
10^{49} erg \cdot s^{-1}$.  This yields a luminosity distance of 55 Mpc
and z = 0.012.  However, Norris, Marani, \& Bonnell (2000) demonstrate
that GRB980425 is a factor of several
hundred below the relation in eq. 2, which suggests that the true
lag/luminosity relation is a broken power law.  If so, then the lag for
GRB830801 implies $L_{lag} = 3 \times 10^{47} erg \cdot s^{-1}$.  This yields 
a luminosity distance of 3.2 Mpc and $z = 0.0007$.  Thus, given that GRB830801
 is by far the brightest known burst and is amongst the lowest luminosity 
events, we know that GRB830801 must be one of closest bursts, perhaps 
substantially closer than even GRB980425 (with SN1998bw).

GRB830801 has $T_{90} > 13 s$, so currently popular ideas suggest that a Type 
Ib or Ic supernova should accompany the burst, with a peak around 15 August
1983.  For a SN1998bw-like event (with $M = -18.88 \pm 0.05$ at peak [Galama 
et al. 1998]), the GRB830801 supernova should have gotten as bright as
14.8 mag (from eq. 1) or 8.6 mag (with a broken power law to accommodate
GRB980425).

On realizing the possibility that GRB830801 might have produced a
supernova visible in binoculars, our first reaction was to check various
supernova catalogs.  For this examination, we used the timing
triangulation position from the InterPlanetary Network of burst detectors
(on Prognoz 9, Vela 5A, Vela 5B, and ISEE) with a position of 11h 58.3m
+11 $^{\circ}$ 50.7' (B1950) and a $1 - \sigma$ uncertainty radius of 
roughly $0.23^{\circ}$.  Out of the supernova catalogs, three known
events (SN1985F, SN1986J, and SN1983ab) were intriguing but ultimately
rejected due to either wrong peak dates or positions (Tsvetkov
1986; Antipin 1996).  However, the burst position was $31^{\circ}$ from 
the Sun when
the supernova would have peaked and was in conjunction with the Sun in the
middle of September.  This can easily explain why no supernova was
discovered near peak.  

But a bright supernova can still be discovered long after peak with
archival plates.  We examined the Harvard College collection of plates,
for which the Damon series covered the position to a median blue
magnitude of 15.2.  No supernova was detected on plates DNB3820 (7
November 1983), DNB3875 (6 December 1983), and DNB3998 (8 February
1984).  The first image is around 94 days after peak, at which time a
SN1998bw-like event will be 3.0 mag below peak (McKenzie \& Schaefer
1999).  So we conclude that any
supernova associated with GRB830801 must have peaked fainter than roughly
12.2 mag.

\section{LUMINOSITIES AND RED SHIFTS}

From Section 2, we have strong confidence in the luminosity indicators, so
we can derive two independent L values for each burst.  In general, the
$L_{lag}$ has a $\sim 2-3$ times smaller uncertainty than the $L_{var}$ 
(based on the scatter about the calibration curves).  However, at high
luminosities, the quantization errors in measuring the lag will
substantially increase the uncertainties in the derived luminosity.
Yet at low luminosities, the variability becomes highly uncertain due to
the normal Poisson noise in the light curve.  We have combined 
$L_{lag}$ and $L_{var}$ as a  weighted average to produce a combined
$L_{c}$ value.  Specifically, we combined the logarithms of the two
luminosity measures where the weights are the inverse square of the
measurement uncertainty as given in section 2.  This luminosity has 
the accuracy of the lag relation at low luminosities, does not suffer from
quantization at high luminosities, and uses all
available information.  The $L_{c}$ values can be combined with the
observed BATSE peak fluxes to derive a luminosity distance (from
eq. 1) and then a red shift ($z$).

In all, we have 112 
GRBs with both luminosities and red shifts.  These are plotted in Figure 2.  
We find $L_{c}$ is between $1.4 \times 10^{50} erg \cdot s^{-1}$ and $2.1
\times 
10^{53} erg \cdot s^{-1}$ with a median of $2 \times 10^{52} erg \cdot 
s^{-1}$; while the red shift varies
between 0.25 and 5.9 with a median of 1.5.  If the calibration curves
are broken power laws as indicated by GRB980425, then the lower limits on
$L_{c}$ and z will be substantially lowered.  Of these bursts, 96 are above our
completeness threshold of $P_{256} > 3.25 photons \cdot s^{-1} 
\cdot cm^{-2}$.  

We have tried to find a signal due to the cosmological dilation of burst
light curve time scales.  With our red shifts, we can divide bursts up
into fairly narrow bins such that burst time scales should vary as
1+z.  We have searched for dilation with three time scales; $T_{90}$, the 
mean peak-to-peak time, and the time from the first-to-last peak.  We have
found no such correlation.  The lack of an apparent dilation effect is
easily understood since our sample was selected to have $T_{90} > 20 s$ in
our rest frame.  This range of $T_{90}$ does not include the peak, so all 
we see is a truncated tail of the distribution.  A truncated tail at one
red shift is little different from a truncated tail at another red shift,
so we should expect little difference.  Also, any comparison of high and
low red shift bursts has the additional complication that the comparison
involves bursts of greatly different luminosity, and there might well be
luminosity/duration correlations.

We have looked for correlations between burst average spectral hardness 
and luminosity.  A hardness/luminosity relation would not suffer from the
definitional problems and the large systematic errors inherent in any
analysis and interpretation of a hardness/intensity relation (Schaefer
1992).  We find that the hardness ratio between BATSE channels 3 and 1 do
change significantly with luminosity in that the luminous bursts are
harder than faint bursts.  To avoid selection effects from BATSE's
trigger, we can isolate those bursts within small ranges of red
shift.  For the 48 bursts from $0.5 < z < 1.5$, the hardness increases
from $3.2 \pm 0.4$ around $10^{51} erg \cdot s^{-1}$ to $5.5 \pm 0.6$
around $2 \times 10^{52} erg \cdot s^{-1}$, while other red shift ranges
have similar shifts.

We find no significant correlation between hardness and red shift, as
might have been expected for cosmological shifting of the peak
energy.  However,
as the low luminosity events must be nearby and the high luminosity events
tend to be very distant, the effect from the previous paragraph will
approximately offset the cosmological shift resulting in the lack of any
apparent correlation.

The luminosities and red shifts displayed in Fig. 2 can be used to derive
the GRB number density ($n_{grb}$) as a function of red shift as well as the
GRB luminosity function.  By taking horizontal strips which do not pass
our completeness threshold of $P_{256} = 3.25 photons \cdot s^{-1} \cdot
cm^{-2}$, the number of
bursts in red shift bins can be divided by the volume to yield a relative
number density.  By taking vertical strips, the number of bursts in
luminosity bins will give the luminosity function.  With both procedures,
the paucity of bursts far from the completeness threshold implies that
any one strip can give only a segment of the desired function, so the
complete function must be pieced together with results from multiple
strips.

Figure 3 displays our derived luminosity function, taken as the number of
bursts appearing within luminosity bins of width $10^{50} erg \cdot
s^{-1}$.  The luminosity
function appears as a broken power law with the break at $2 \times 10^{52} 
erg \cdot s^{-1}$.
This luminosity break does not correspond to the possible breaks in the
lag and variability relations suggested on the basis of GRB980425.  The
dependence above the break is fitted to be scaling as $L^{-2.8 \pm 0.2}$, 
while it scales as $L^{-1.7 \pm 0.1}$ below the break.

Figure 4 displays our resulting $n_{grb}$ as a function of $z$.  The power
law dependence is roughly $(1+z)^{2.5 \pm 0.3}$ for $0.2 < z < 5$.  This
result clearly
rejects scenarios for which $n_{grb}$ does not evolve with
distance.  For $z < 2$, our
result is easily consistent with the burst number density varying as
the star formation rate (Steidel et al. 1999), as is expected since long
duration GRBs are formed from the deaths of massive stars.  That is,
$n_{grb}$ should closely follow the star formation rate in our Universe.

However, it is surprising and exciting that $n_{grb}$ keeps rising
monotonically from $2<z<5$.  The surprise is because the star formation
rate is widely taken to either to be flat or to fall substantially above a
red shift of $\sim 2$ (Steidel et al. 1999).  But all previous measures
have had major problems with reddening at high red shift.  Gamma radiation
is not affected by reddening and thus the star formation rate in Fig. 4
might be the first view of the true situation.  

        We thank Michael Boer, Kevin Hurley, and John Laros for communication 
of their IPN position of GRB 830801.  Martha Hazen and Alison Doane helped
with the Harvard plate collection.

\begin{figure}
\begin{center}
\resizebox{8cm}{9cm}{\includegraphics{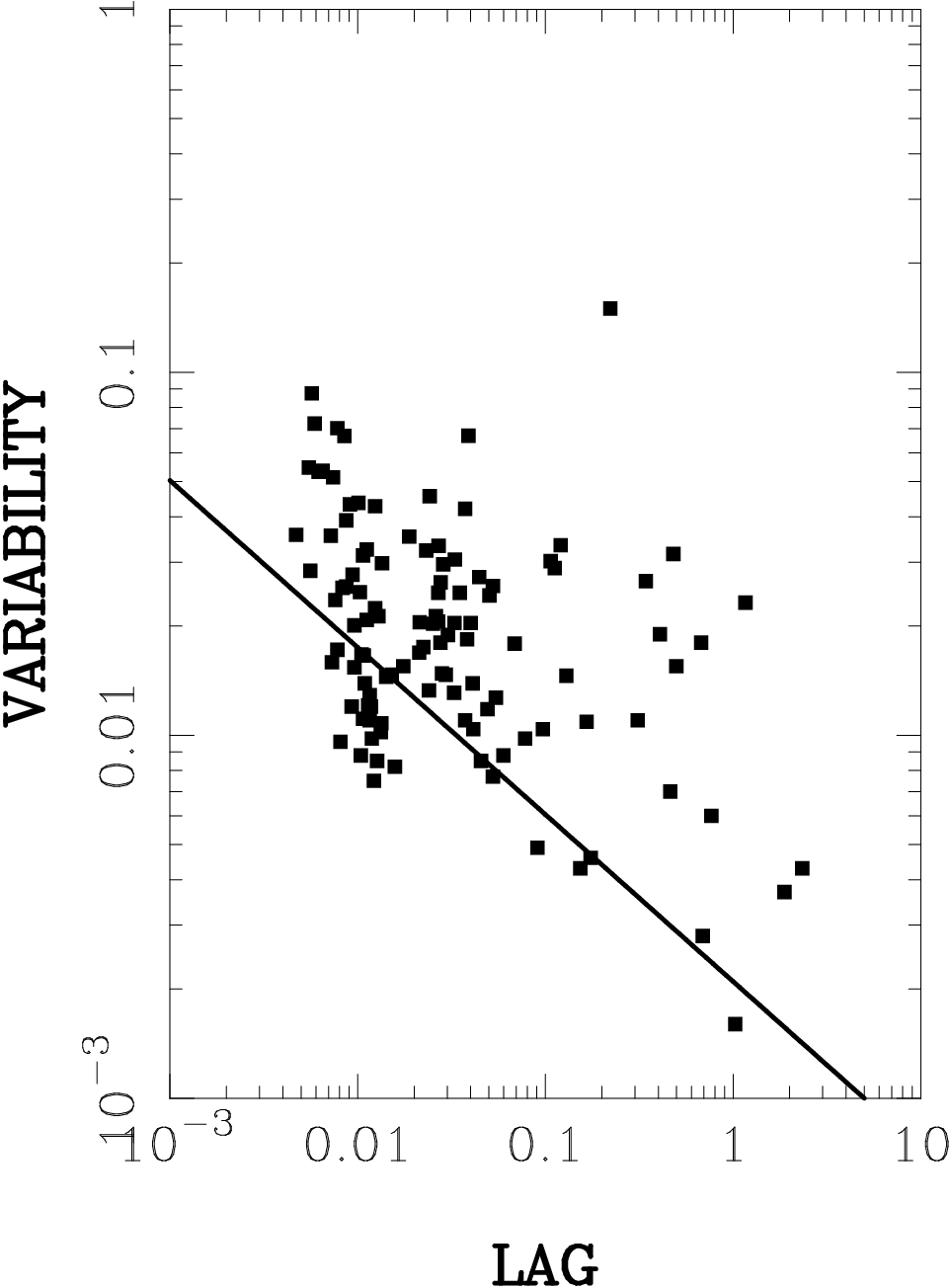}}
\caption{The lag/variability correlation for 112 BATSE bursts.
If both the lag/luminosity and the variability/luminosity relations are
true, then there must be a lag/variability relation shown by the straight
line.  Indeed, the correlation coefficient is $r = -0.45$ which shows a
correlation at the 99.999924\% confidence level, while the best fit line 
has the predicted slope.  The intercept is roughly a factor of two low,
but this is well within the uncertainties.  The successes of the predicted
correlation (its existence, and slope) prove that {\em both} luminosity 
indicators are valid.  The plotted values have red shift effects
removed.  The measured lags are quantized to 0.064 seconds, and the
$\tau_{lag}$ values measured as zero are
displayed as if $\tau_{lag} = 0.032$ for this logarithmic scale.}
\end{center}
\end{figure}

\clearpage
\begin{figure}
\begin{center}
\resizebox{8cm}{9cm}{\includegraphics{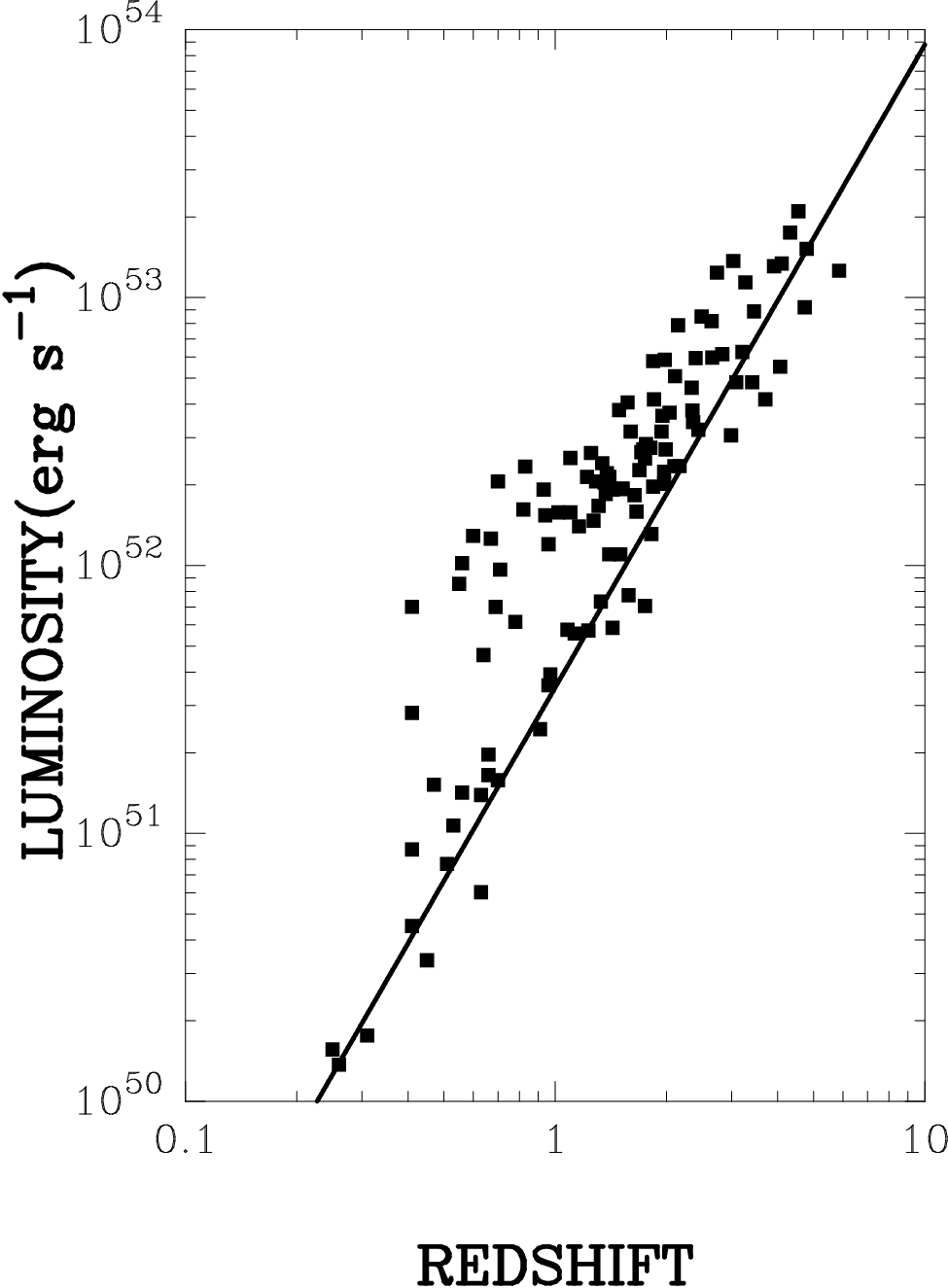}}
\caption{The luminosity and red shift of 112 BATSE bursts.
The burst luminosities were derived from a weighted average of the two
luminosity indicators.  The red shifts were derived from the luminosities
and the measured peak fluxes.  The diagonal line is our line of
completeness at $P_{256} = 3.25 photons \cdot cm^{-2} \cdot s^{-1}$.  
Cuts in the vertical direction can give the burst luminosity function (see
Fig. 3).  Cuts in the horizontal direction can give the number density of
bursts as a function of red shift (see Fig. 4).  So for example, a 
horizontal strip around a luminosity of $10^{53} erg \cdot s^{-1}$ shows
that bursts with $z>2$ have a higher rate than bursts with $z<2$.}
\end{center}
\end{figure}

\clearpage
\begin{figure}
\begin{center}
\resizebox{8cm}{9cm}{\includegraphics{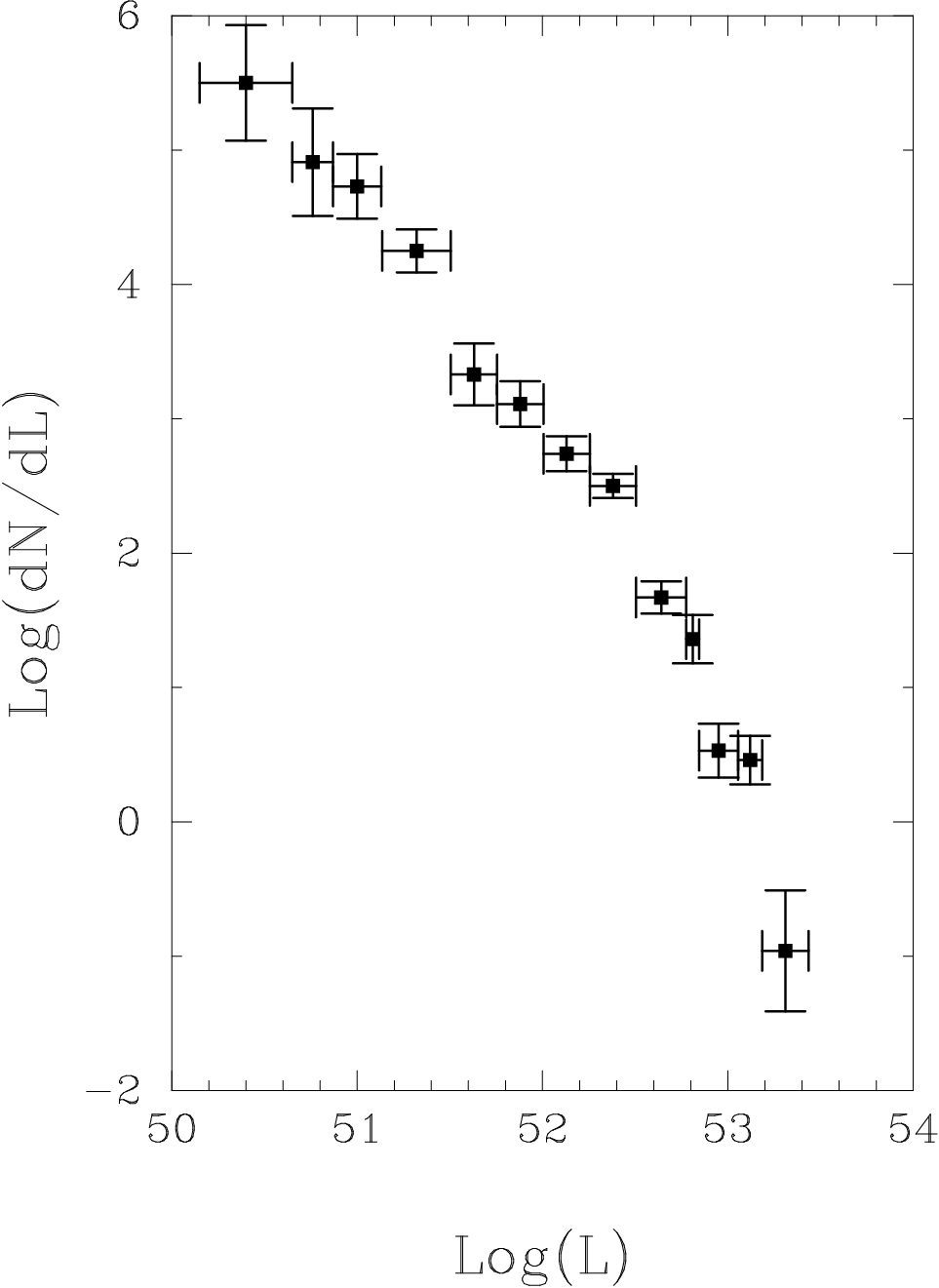}}
\caption{The GRB luminosity function.  This measured luminosity function shows
the (arbitrarily scaled) number of bursts that appear within a luminosity 
bin with width $10^{50} erg \cdot s^{-1}$
as a function of luminosity.  The functional form is a broken power
law.  This luminosity function now opens the the possibility of many
exciting demographic studies of GRBs.} 
\end{center}
\end{figure}

\clearpage
\begin{figure}
\begin{center}
\resizebox{8cm}{9cm}{\includegraphics{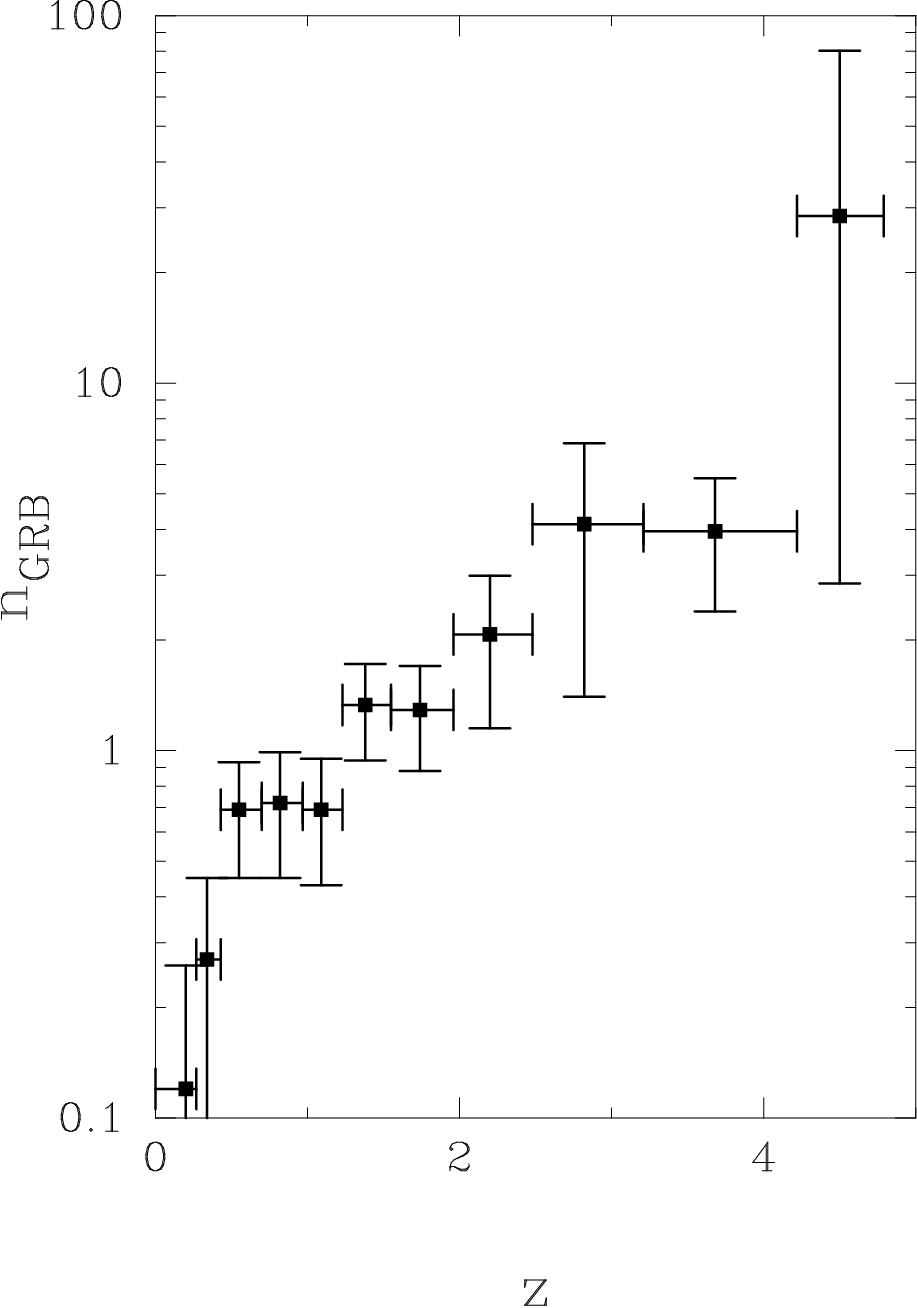}}
\caption{The rate evolution of bursts with distance.
The measured number density of bursts varies as $(1+z)^{2.5 \pm 0.3}$ from
red
shifts 0.2 to 5.  This provides two important results.  First, at low red
shifts, $n_{grb}$ follows the well-measured star formation rate, as
expected since long duration GRBs are created by the deaths of massive
stars.  Second, at high red shifts, $n_{grb}$ and hence the star formation
rate of our Universe are rising monotonically out to $z \sim 5$.  Given
that gamma radiation is not affected by reddening, the majority of our
Universe's star formation occurred substantially earlier than previously
realized.} 
\end{center} 
\end{figure}

\end{document}